\documentstyle[aps,prl,multicol,epsf,rotate]{revtex}
\draft

\begin{document}

\title{The effect of an in-plane magnetic field on the interlayer 
transport of quasiparticles in layered superconductors}

\author{L.~N.~Bulaevskii, M~J.~Graf, and M.~P.~Maley}
\address{Los Alamos National Laboratory, Los Alamos, NM 87545}

\date{received: January 29, 1999}

\maketitle
\begin{abstract}
We consider the quasiparticle $c$-axis conductivity in highly 
anisotropic layered compounds in the presence of the magnetic field 
parallel to the layers. We show that at low temperatures the 
quasiparticle interlayer conductivity depends strongly on the 
orientation of the in-plane magnetic field if the excitation gap has 
nodes on the Fermi surface. Thus measurements of the angle-dependent 
$c$-axis (out-of-plane) magnetoresistance, as a function of the 
orientation of the magnetic field in the layers, provide information 
on the momentum dependence of the superconducting gap (or pseudogap) 
on the Fermi surface.  
Clean and highly anisotropic layered superconductors seem to be the 
best candidates for probing the existence and location of the nodes 
on the Fermi surface.
\end{abstract}

\pacs{PACS: 74.25.Fy, 74.70.Kn, 74.72.-h}

\begin{multicols}{2}

The symmetry of the excitation gap in the superconducting and normal
state of the cuprates, the quasi-two-dimensional organic salts, and the
ruthenate oxides has been the focus of theoretical and experimental
studies for the last several years.
Now there is consensus that most of the cuprates are not only
anomalous metals that develop a pseudogap when underdoped
\cite{arpes}, but also that they are unconventional ($d$-wave)
superconductors 
with nodes on the Fermi surface, as demonstrated in phase-sensitive
Josephson junction experiments \cite{Kirt}.
The situation is less clear in the organic salts and the ruthenates,
in spite of many reports of power laws in the temperature behavior of
various transport and thermodynamic properties (see, e.g.,
Refs.~\onlinecite{kanoda,rev}).  Spin fluctuation models for the
organics predict a $d_{xy}$ superconducting state 
similar to the cuprates\cite{rev,sch}. Since phase-sensitive
Josephson junction and angle-resolved photoemission spectroscopy
experiments are not available for 
these materials, other more stringent experiments are required.

In this Letter we propose angle-dependent magnetoresistance oscillation
(AMRO) experiments which directly probe the locations of the line
nodes of the gap in the quasiparticle excitation spectrum 
in the superconducting phase or in the normal state with a pseudogap. 
We show that in
highly anisotropic layered materials the dependence of the 
quasiparticle $c$-axis I-V characteristics on the orientation 
of a magnetic field parallel
to the layers, ${\bf B}_{\parallel}$, allows one to extract 
the momentum dependence of the gap function. 
In particular, we calculate the quasiparticle $c$-axis conductivity, 
$\sigma_q$, in Josephson coupled layered gapless superconductors 
(e.g., $d$-wave or $p$-wave pairing) as a function of 
${\bf B}_{\parallel} = (B_x, B_y)$ and compare it with results for 
$s$-wave pairing. 
We assume a significant contribution of coherent electron tunneling 
between adjacent layers.  
The condition for coherent tunneling in the presence of an in-plane
magnetic field requires 
$|\delta {\bf k}| \ll |{\bf b}|$, where $|\delta {\bf k}|$ 
is the change of the in-plane momentum of a tunneling electron, 
where ${\bf b}=(2\pi s/\Phi_0){(B_y, -B_x)}$, and $s$ is the 
interlayer spacing.
We show that under such conditions the dependence of $\sigma_{q}$ on 
the orientation of ${\bf b}$ enables one to determine whether the 
superconducting gap (or any other kind of gap) has nodes on the Fermi 
surface, as well as information on the location of these nodes.
The conditions that are necessary to apply our method seem to be 
fulfilled in highly anisotropic and very clean organic salts like the
quasi-two-dimensional BEDT-TTF [bis(ethylenedithio)-tetrathiafulvalene]
superconductors \cite{rev}, the Bi- or Tl-based cuprates 
\cite{lat,coop}, and the oxide superconductor Sr$_2$RuO$_4$ 
\cite{ohmichi}.

\begin{figure}[h]
\begin{minipage}{0.48\textwidth}
\epsfxsize=1.60in
\centerline{\rotate[r]{\epsfbox{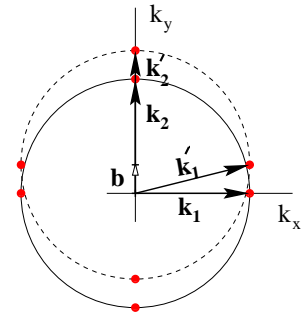}}}
\vspace{4pt}
\caption {
Sketch of the shifted Fermi circles in adjacent layers, which a 
tunneling electron sees, due to an applied
in-plane magnetic field ${\bf b}$.  The filled circles indicate the
nodal regions of a $d_{xy}$ order parameter with nodal vectors
$|{\bf k}_1| \approx |{\bf k}_1'|$ and 
$|{\bf k}_2| = |{\bf k}_2'| - |{\bf b}|$.
}\label{cartoon}
\end{minipage}
\end{figure}

For perfect crystals with translational invariance 
we use the interlayer Hamiltonian, which describes the conservation of 
the in-plane momentum when electrons tunnel between the layers.
In Josephson coupled 
superconducting layers the in-plane magnetic field penetrates almost 
freely into the sample, inducing the vector potential 
$a_z({\bf r})= {\bf b \cdot r}$, where ${\bf r}=(x,y)$ 
is the in-plane coordinate. Thus the interlayer Hamiltonian or tunneling
Hamiltonian can be written as
\begin{eqnarray}
&& {\cal H}_{\perp} = 
t^{}_{\perp}\! \sum_{n,\sigma}\int\!d^2{\bf r}
\big[\psi^{+}_{n,\sigma}({\bf r})
 \psi^{}_{n+1,\sigma}({\bf r})
 e^{i\chi^{}_{n,n+1}({\bf r})}+h.c.
\big] ,
\nonumber \\
&&\chi^{}_{n,n+1}({\bf r})=\frac{1}{s} \int_{ns}^{(n+1)s}\! {dz}
\, a_z({\bf r}) ,
\end{eqnarray}
where $\psi^{}_{n,\sigma}$ is the annihilation operator for electrons 
in layer $n$ with spin $\sigma$, and $t^{}_{\perp}$ is the interlayer 
transfer integral.  We assume that $t^{}_\perp$ is isotropic in the 
layers. 
The vector potential $a_z({\bf r})$ leads to a change in the in-plane 
momentum of the tunneling electron, ${\bf k} \to {\bf k} + {\bf b}$, 
where ${\bf k}=(k_x,k_y)$ is the in-plane momentum. 
In the momentum representation the interlayer tunneling 
Hamiltonian has the form 
\begin{equation}\label{ck} 
{\cal H}_{\perp}({\bf b})=
t^{}_{\perp}\! \sum_{n,{\bf k},\sigma}
\big[ \psi^{+}_{n,{\bf k},\sigma}
  \psi^{}_{n+1,{\bf k}+{\bf b},\sigma}
  + h.c.
\big].
\end{equation}
Defects in the crystal structure lead to spatial variations in 
$t_{\perp}$. The spatial average, $\langle t_{\perp}\rangle$, 
determines the coherent tunneling part. 
We assume that it dominates the low-temperature $c$-axis transport and 
use in (1) and (2) $t_\perp = \langle t_\perp \rangle$, for details see 
Ref.~\onlinecite{lat}.
Scattering inside the layers leads to a change of the momentum 
while tunneling,
$|\delta {\bf k}|=1/\ell_{\parallel}$. 
Here $\ell_{\parallel}$ is the effective mean-free-path for 
scattering of quasiparticles  inside the layers. 
In conductivity measurements a finite voltage $V$ is applied between 
neighboring 
layers and hence the energy of the tunneling electron changes by $V$. 
In the limit $V\to 0$ the energy is conserved within the accuracy of 
the temperature $T$. The important point is that 
at low temperatures only quasiparticles near the nodes of the gap
on the Fermi surface can contribute to the dissipative 
quasiparticle $c$-axis transport. 
Let us denote the positions of the nodes on the Fermi surface by the 
momenta ${\bf k}_g$. Then at low $T$ and $b\gg 1/\ell_{\parallel}$ 
the contribution of 
the quasiparticles near the nodes to the conductivity is 
significant when the change of the quasiparticle energy, 
$E_{{\bf k}+{\bf b}}-E_{{\bf k}}$, is small, i.e.,
when $\hbar{\bf v}_f({\hat{\bf k}}_g) \cdot {\bf b}\ll T\ll \Delta_0$,
see Fig.~\ref{cartoon}.
Here $\hat{{\bf k}}={\bf k}/k$ is a unit vector, ${\bf v}_f$ is the 
Fermi velocity, and $\Delta_0$ is the amplitude of the energy gap. 
This means that the dominant contribution to 
$\sigma_{q}$ comes from quasiparticles in the vicinity of the nodes
and contribution of a given node ${\bf k}_g$ to $\sigma_q$ is maximal 
for ${\bf b} \perp {\bf v}_f(\hat{\bf k}_g)$.


In the superconducting state the Josephson current
contributes to the $c$-axis transport.
Thus to obtain only the quasiparticle contribution the 
Cooper pair current must be suppressed when the interlayer
current is measured as a function of the orientation of
the in-plane magnetic field.  The suppression of the Josephson 
current can be achieved by applying a $c$-axis current significantly 
exceeding the Josephson critical current \cite{suz}. 
Such a condition is easily fulfilled, considering that the in-plane 
magnetic field strongly diminishes the Josephson critical current, 
without significantly affecting the superconductivity inside the layers.
Another method for observing the quasiparticle current is to perform the
measurements of the $c$-axis conductivity in the resistive state of 
the intrinsic Josephson junctions \cite{klei}. 

We calculate the quasiparticle conductivity $\sigma_{q}$ using the 
BCS theory for $d$-wave and $p$-wave pairing in the presence of elastic,
isotropic in-plane scattering, characterized by the normal-state
scattering rate $\Gamma \ll\Delta_0$. 
Here $\Delta_0$ is the amplitude of the 
superconducting spin-singlet $d$-wave gap 
[$\Delta(\hat{{\bf k}})=\Delta_0(T)(\hat{k}_x^2-
\hat{k}_y^2)$ or $\Delta(\hat{{\bf k}})=2\Delta_0(T)\hat{k}_x\hat{k}_y$]
or of the spin-triplet $p$-wave gap along a given spin direction
[$\Delta(\hat{\bf k})=\Delta_0(T)\hat{k}_x$ or 
$\Delta(\hat{\bf k})=\Delta_0(T)\hat{k}_y$]. 
We consider first a circular Fermi surface inside 
the layers and a cylindrical, three-dimensional open Fermi surface with 
$t^{}_{\perp}$ much smaller than any other relevant energy scale in 
the system.  For simplicity, we neglect the effect of the Zeeman term 
on the in-plane Green functions \cite{yang}. Although it will change 
the field dependence of $\sigma_q$, it will not affect its angular 
dependence.  The perturbation theory with respect to ${\cal H}_{\perp}$ 
gives the expression for the quasiparticle conductivity
in terms of the quasiparticle spectral functions \cite{mahan}:
\begin{eqnarray}
\sigma_{q}({\bf b}) & = & \frac{e^2 t^{2}_{\perp} s}{4 \pi T \hbar}
\int_{-\infty}^{+\infty}\!\!d\omega \,
\cosh^{-2}\frac{\omega}{2 T}
\nonumber \\ && \qquad \times
\int\!d^2{\bf k}\, A({\bf k}+{\bf b},\omega) A({\bf k}, \omega) , 
\label{gr}
\end{eqnarray}
where ${\rm Im}G({\bf k},\omega)= -\pi A({\bf k},\omega)$ is the 
imaginary part of the Green function.  Within the BCS theory we obtain  
$$
A({\bf k},\omega)=
\frac{(1+\xi_{{\bf k}}/E_{{\bf k}})\gamma}{
 2\pi[(\omega'-E_{{\bf k}})^2+\gamma^2] } +
\frac{(1-\xi_{{\bf k}}/E_{{\bf k}})\gamma}{
 2\pi[(\omega'+E_{{\bf k}})^2+\gamma^2] } \, ,
$$
where $E_{{\bf k}}=[\xi_{{\bf k}}^2+\Delta^2(\hat{{\bf k}})]^{1/2}$, and
$\xi_{{\bf k}}$ is the quasiparticle energy in the normal state,
measured from the Fermi level. The scattering rate of the 
quasiparticles is $\gamma(\omega)=-2{\rm Im}\Sigma(\omega)$, where 
$\Sigma(\omega)$ is 
the diagonal part proportional to the unit matrix of the impurity 
self-energy, and $\omega'(\omega)=\omega-{\rm Re} \Sigma(\omega)$. 
The impurity self-energy has to be calculated self-consistently
and is energy (temperature) dependent, even for elastic 
scattering, in the superconducting state \cite{pet}.

In unconventional superconductors even nonmagnetic impurities lead to 
gapless states in the vicinity of the nodes of the gap function on the 
Fermi surface.
At low temperatures, $T \ll \gamma(0) \ll \Delta_0(0)$,
the dominant contribution to the quasiparticle conductivity comes from 
these gapless states.  In rather clean superconductors and at low 
energies $\gamma(\omega)$ becomes
$\gamma(0)\sim [\Gamma\Delta_0(0)]^{1/2}$ in the 
case of strong scattering, and 
$\gamma(0)\sim \Delta_0(0)\exp[-\Delta_0(0)/\Gamma]$ 
for weak scattering (Born limit).
Note that the relevant phase space for scattering of quasiparticles
is restricted to a fraction $\sim \gamma(0)/\Delta_0(0)$ of phase
space.
At low temperatures, $T \ll \gamma(0)$, we obtain
\begin{equation}
\sigma_{q}({\bf b}) \approx \frac{e^2 t^{2}_{\perp} s}{\pi^3 \hbar}
\int\!d^2{\bf k} \,
A({\bf k}+{\bf b},0) A({\bf k},0).
\label{gr0}
\end{equation}
In the following, we
take into account that regions near the nodes contribute mainly to 
the integral over ${\bf k}$.  We replace 
$E^2_{{\bf k}+{\bf b}}\approx \xi^2_{{\bf k}+{\bf b}} 
 + \Delta^2(\hat{\bf k})$,
where
$\xi_{{\bf k}+{\bf b}} \approx \xi_{{\bf k}}+ \hbar {\bf b}\cdot 
{\bf v}_f(\hat{{\bf k}})$, and parameterize
$\hat {\bf k}=(\cos\varphi, \sin\varphi)$ and
${\bf b}=b(\cos\theta,\sin\theta)$. 
For example, the $N_g=4$ nodal angles for d$_{x^2-y^2}$ pairing are 
given by $\varphi_g=(2 g - 1)\pi/4$.
Finally, we find at low temperatures and
at low fields a universal conductivity, similar to the in-plane 
transport \cite{universal},
\begin{eqnarray}
&&\sigma_{q}(b=0) \approx
\frac{e^2 t^{2}_{\perp} s N_f}{\hbar} \frac{2}{\pi \Delta_0(0)} =
\frac{ e s J_0}{\pi \Delta_0(0)},
\label{s0} \\
&&\frac{\sigma_q(\theta)}{\sigma_q(b=0)} \approx
\frac{1}{N_g}
\sum_{g=1}^{N_g}
\frac{ \ln \big[ ({1+\alpha_g^2})^{1/2} + |\alpha_g| \big] }{
 |\alpha_g| ({1+\alpha_g^2})^{1/2}} \,,
\label{gr2}
\end{eqnarray}
with $\alpha_g=(\ell_{\|}/2 v_f) {\bf b}\cdot{\bf v}_f(\hat{\bf k}_g)
= (\ell_{\parallel} b/2)\cos(\varphi_g-\theta)$, where
$\ell_{\parallel}= \hbar v_f/\gamma(0)$ is the effective scattering 
mean-free-path, and $J_0 =2 e t^2_\perp N_f / \hbar$ is the 
Josephson critical current density.
Here $N_f = m / (2\pi^2 \hbar^2)$ is the 2D density of states per spin
of quasiparticles with mass $m$. 
This result is readily generalized to an elliptical 2D Fermi surface by 
rescaling the momenta $k_i = \sqrt{m_i/m}\, k_i'$, ($i=x,y$),
where $m_i$ are the effective masses and $m^2 = m_x m_y$. Then
Eq.~(\ref{gr2}) is valid with $\alpha_g=(\ell_{\parallel}b/2)\big[ 
 \sqrt{m/m_x}\cos\theta \cos\varphi_g
 +\sqrt{m/m_y}\sin\theta \sin\varphi_g \big]$.

\begin{figure}[h]
\begin{minipage}{0.48\textwidth}
\epsfxsize=2.60in
\centerline{\rotate[r]{\epsfbox{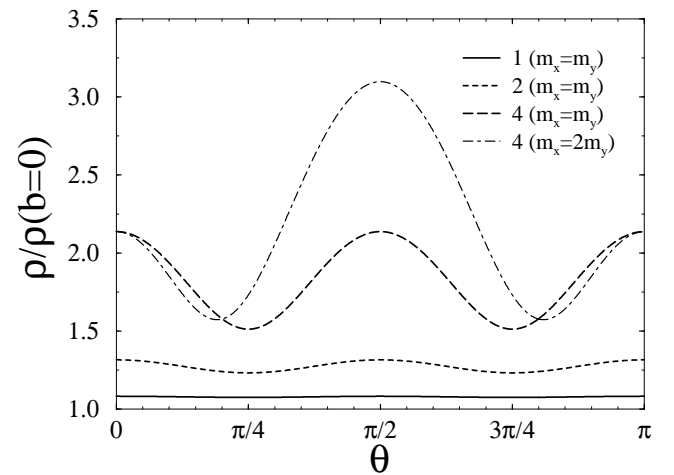}}}
\caption {
AMRO, $\rho=1/\sigma_q$,
at low temperatures, $T \ll \gamma(0)$, for a $d_{x^2-y^2}$-wave
superconductor at different parameters $\ell_\| b = 1, 2, 4$ and
for an elliptical Fermi surface with specified mass anisotropies.}
\label{conductivity}
\end{minipage}
\end{figure}

The largest contribution of a given node to the conductivity comes when 
${\bf b} \perp {\bf v}_f(\hat{\bf k}_g)$. 
The conductivity is maximal for ${\bf b}$ along the nodes with 
periodicity $\theta=\pi/2$, as shown in Fig.~\ref{conductivity}. 
For low fields, $(\ell_\| b/2)^2 \ll 1$), 
$\sigma_q({\bf b})$ shows a quadratic field
dependence.  In the high field limit, $(\ell_\| b/2)^2 \gg 1$, the 
conductivity falls off as $b^{-2} \ln\, b$. 


In the case of $\gamma(0)\ll T\ll\Delta_0(0)$, 
the quasiparticle scattering rate is energy (temperature) dependent 
and $\gamma(T)\sim \Gamma (T/\Delta_0)^n$ with 
$n=1$ in the weak scattering regime (Born limit), 
and $n=-1$ for strong scattering \cite{pet}. 
Thus for fields $b\ll \Delta_0(0)/\hbar v_f$ 
we obtain after integrating over $\omega$
\begin{equation}
\sigma_{q}({\bf b}) \approx \frac{e^2 t^{2}_{\perp} s}{2 \pi^3 T \hbar}
\int\!\!d^2{\bf k}\,
 \frac{\gamma(T)\, \cosh^{-2}({E_{{\bf k}}}/{2T})
}{(E_{{\bf k}+{\bf b}}-E_{{\bf k}})^2+4\gamma^2(T)} \, .
\label{s1}
\end{equation}
In the next step, we expand
$E_{{\bf k}+{\bf b}}-E_{{\bf k}}\approx (\partial E_{{\bf k}}
/\partial {\bf k})\cdot{\bf b}=
E_{{\bf k}}^{-1}\xi_{{\bf k}}\hbar{\bf v}_f(\hat{{\bf k}})\cdot{\bf b}$,
and take into account that at low temperatures, 
$T \ll \Delta_0(0)$, only small values of $\xi_{\bf k}$ 
and angles near the nodes, $\varphi_g$, 
contribute to the quasiparticle conductivity. Then we obtain
\begin{equation}
\frac{\sigma_q(\theta)}{\sigma_q(b=0)} \approx
\frac{1}{N_g}
\sum_{g=1}^{N_g}
\frac{1}{( 1+\alpha_g^2 )^{1/2}},
\label{gr22}
\end{equation} 
where now $\ell_\| \sim v_f \Delta_0/\Gamma T$ in the Born limit 
and $\ell_\| \sim v_f T/\Gamma \Delta_0$ in the strong scattering 
limit. In the high field limit, $(\ell_\| b/2)^2 \gg 1$,
the quasiparticle $c$-axis conductivity falls off as $1/b$. 

These results are quite general and also valid for $p$-wave 
superconductors. In this case $\varphi_g=0,\pi$ or 
$\varphi_g={\pi}/{2}, {3\pi}/{2}$. Now 
$\sigma_{q}(\theta)$ is minimal for fields along 
the nodes with periodicity $\pi$.  
A $p$-wave state has larger oscillations than 
a $d$-wave state for the same parameter $\ell_\| b$.

Angle-dependent oscillations of $\sigma_q(\theta)$ were 
observed in the normal state in organic conductors 
\cite{leb}, cuprates \cite{coop}, and ruthenates \cite{ohmichi},
and were explained in terms of Fermi surface anisotropy.
In materials with an elliptical 2D Fermi surface the angular periodicity
of $\sigma_q(\theta)$ is $\pi$.  The amplitude of the oscillations is
determined by the parameter $\delta v_f b / \Gamma$, where $\delta v_f$ is the 
variation of the anisotropic Fermi velocity.

The periodicity and height of the AMRO, induced by the topology of 
the Fermi surface in the normal state, changes when a
superconducting gap or pseudogap emerges. 
If an orthorhombic (organic) crystal undergoes a superconducting 
transition with $s$-wave pairing, the $\pi$-periodicity will
preserve, but $\sigma_q$ will tend to 
zero as $T\rightarrow 0$. For a $d$-wave state an additional 
$\pi/2$-periodicity emerges on cooling and $\sigma_q$ saturates as 
$T\rightarrow 0$. 
In tetragonal cuprate crystals the periodicity 
is $\pi/2$, both in the superconducting and normal state \cite{coop}. 
The positions of the AMRO maxima in the normal and superconducting 
states may differ (as in Tl-2201 \cite{Kirt,coop}) or coincide. 
In the former case, the maxima observed in the normal state should 
drop and new ones should appear on cooling below $T_c$ and further
increase with increasing field $b$.
In the latter case, the positions of the AMRO maxima do not change 
below $T_c$, but $\sigma_q$ saturates on cooling at $T\rightarrow 0$
for $d$-wave superconductivity,
while for s-wave pairing $\sigma_q$ vanishes exponentially.
The main parameter which determines the angular behavior of
$\sigma_{q}(\theta)$ at low temperatures,
$T\ll T_c$, is $\ell_\| b=v_fb/\gamma$.
We note that $\gamma\gg\Gamma$, while $\delta v_f$ can be of order
$v_f$. Thus, we anticipate that the amplitude of oscillations will 
be about the same above and
below $T_c$ in tetragonal crystals and weaker in orthorhombic crystals.

Using typical values for the
interlayer spacing, $s\approx 15\, {\rm \AA}$,  in organic and cuprate
superconductors, e.g.,  Bi$_2$Sr$_2$CaCu$_2$O$_{8-\delta}$ (Bi-2212) 
crystals, we obtain $b\approx 4.56\, B \,{\rm (T\, \mu m)^{-1}}$.
Then at low temperatures a mean-free-path of 
$\ell_\| \gtrsim 200\, {\rm \AA}$ 
is needed for a  $d$-wave state in order to observe significant
oscillations, i.e., oscillations that are larger than $1\%$
of the total magnetoresistivity for magnetic fields of order 
$10\, {\rm T}$. It is feasible that this condition is fulfilled 
in the organic BEDT-TTF superconductors, 
where $\ell_{\parallel} \approx 1000\, {\rm\AA}$ was deduced from 
de Haas-van Alphen and Shubnikov-de Haas measurements \cite{org}.
In the high-temperature superconductors Bi-2201 and Bi-2212 the 
mean-free-path, $\ell_\|$, of quasiparticles at low temperatures is 
sufficiently long \cite{lat,coop} to observe this effect, too. 

Until now we have discussed the dependence of the quasiparticle 
conductivity 
on the orientation of ${\bf b}$. More generally, 
the dependence of $\sigma_q$ on ${\bf b}$, described by 
Eq.~(\ref{gr}), provides information on the spatial correlations in the 
system. Namely, if $\sigma_q({\bf b})$ is obtained experimentally then 
the spatial dependence of the spectral density of the Green function 
may be extracted by an inverse Fourier transformation with respect 
to ${\bf b}$,
\begin{eqnarray}
&&(1/4T)\int_{-\infty}^{+\infty}\!\!\!\!d\omega\, 
 \cosh^{-2}\frac{\omega}{2 T}\,
 A({\bf r},\omega) A(-{\bf r}, \omega) =  \qquad\qquad \mbox{}
\nonumber \\ && \qquad\qquad\qquad
(\pi^3\hbar/e^2 t^{2}_{\perp} s)
\int\!d^2{\bf b}\, e^{i{\bf b}\cdot{\bf r}}
\sigma_{q}({\bf b}) . 
\end{eqnarray}
In the low temperature regime the left-hand side is simply 
$A({\bf r},0) A(-{\bf r},0) = A^2({\bf r},0)$ and the coordinate 
dependence of the spectral function at 
zero energy may be calculated directly from measurements of 
$\sigma_q({\bf b})$. 
A similar method was used previously to obtain information on the 
correlation function of pancake vortices in a vortex liquid from the 
dependence of the plasma resonance frequency or from the dependence 
of $\sigma_{\perp}$ on the magnetic field parallel to the layers with 
a fixed out-of-plane component \cite{kosh,mor}. 
For superconductors, in the absence of a perpendicular magnetic field,
the coordinate dependence of $A({\bf r},0)$
provides information on the superconducting correlation length.
At low temperatures in highly anisotropic layered metals in the normal 
state the combination of parallel and perpendicular magnetic field 
components leads to oscillations of the $c$-axis conductivity as a 
function of $b$ or as a function of the angle between the magnetic 
field and the $c$-axis \cite{McK,yak}. 

The coordinate dependence of the spectral density of the Green function
at arbitrary energy may be extracted from $c$-axis I-V measurements in
the presence of an in-plane magnetic field. The $c$-axis current density
is 
\begin{eqnarray}
&&J_{q}({\bf b},V)=
\frac{e t^{2}_{\perp} s}{2 \pi^3 \hbar}
\int_{-\infty}^{+\infty}\!\!\!\!d\omega\,
\left[\tanh\frac{\omega+e V}{2T}-\tanh\frac{\omega}{2T}\right]
\nonumber \\ &&\qquad\qquad \times 
\int\!\!d^2{\bf k}\,
A({\bf k}+{\bf b},\omega+e V) A({\bf k}, \omega). 
\label{g}
\end{eqnarray}
In the limit of low 
temperatures, $T \to 0$, the spectral densities are only slowly varying
with respect to $\omega$.  In this limit we
solve the integral equation (\ref{g}) for $A({\bf r},\omega)$ at 
known $J_q({\bf b},V)$ by Fourier and Laplace transformation,
\begin{eqnarray}
&&\int_{0}^{\infty}\!\!\!\!d\omega\, e^{-\omega \tau} A({\bf r}, \omega)
 = [{2 \pi^3 \hbar}/{t^{2}_{\perp} s}]^{1\over 2}
\nonumber\\ &&\qquad\qquad \times
\left[
  \int_{0}^{\infty}\!\!\!\!dV\, e^{-e V \tau}
  \int\!\!d^2{\bf b}\, e^{i{\bf b}\cdot{\bf r}} J_{q}({\bf b},V)
\right]^{1\over 2} \,.
\end{eqnarray}

In conclusion, we have shown that, in principle, at low temperatures 
the spatial dependence of the spectral function $A({\bf r},\omega)$ 
can be determined from measurements of the $c$-axis I-V characteristic 
as a function of the magnetic field ${\bf B}_{\parallel}$.  
The dependence of $\sigma_q$ on the orientation of the in-plane 
field ${\bf B}_\parallel$ provides information on the existence of 
line nodes on the Fermi surface as well as on their locations.  
Measurements of the angle-dependent in-plane
magnetoresistance will be useful to map out the momentum distribution of
the superconducting gap or pseudogap in organic salts and cuprates.

We thank V.M.\ Yakovenko and A.V.\ Balatsky for many useful discussions.
This work was supported by the Los Alamos National
Laboratory under the auspices of the U.S.\ Department of Energy.


%
%

\noindent

\end{multicols}
\end{document}